\documentclass[a4paper,11pt]{article}

\usepackage{latexsym}
\usepackage{amsmath}
\usepackage{amssymb}
\usepackage{amsfonts}
\usepackage{amsthm}
\usepackage[latin1]{inputenc}
\usepackage{graphicx,subfigure}
\usepackage{float,times}
\usepackage[usenames]{color}
\usepackage{longtable}
\usepackage{rotating}
\usepackage[round]{natbib}
\usepackage{fancyhdr}
\usepackage{multirow}
\usepackage[margin=2cm]{geometry}
\usepackage[center]{titlesec}

\newtheoremstyle{example}{\topsep}{\topsep}%
     {}
     {}
     {\bfseries}
     {}
     {\newline}
     {\thmname{#1}\thmnumber{ #2}\thmnote{ #3}}

\theoremstyle{example}
\theoremstyle{theorem}
\theoremstyle{theorem}
\theoremstyle{proposition}
\theoremstyle{corollary}

\def\keywords{\vspace{.3cm}
{\textit{Keywords}:\,\relax%
}}
\def\endkeywords{\par}

\title{Bayesian Survival Modelling of University Outcomes}

\author{Catalina A. Vallejos$^{1,2}$  and   Mark F.J. Steel$^3$\thanks{Corresponding author: Mark Steel, Department of Statistics, University of Warwick, Coventry, CV4 7AL, UK; email: M.F.Steel@stats.warwick.ac.uk. During this research, Catalina Vallejos was a PhD student at the University of Warwick. She acknowledges funding from the University of Warwick and the Pontificia Universidad Cat\'olica de Chile (PUC). We are grateful to the PUC for access to the dataset analyzed in this article. We also thank Lorena Correa and Prof.~Guillermo Marshall for motivating the analysis and Valeria Leiva for support in  accessing the data.}\\ $^1$MRC Biostatistics Unit $^2$EMBL-EBI $^3$Dept. of Statistics, Univ. of Warwick}

\date{}

\begin{document}

\maketitle

\begin{abstract}
The aim of this paper is to model the length of registration at university and its associated academic outcome for undergraduate students at the Pontificia Universidad Cat\'olica de Chile. Survival time is defined as the time until the end of the enrollment period, which can relate to different reasons - graduation or two types of dropout - that are driven by different processes. Hence, a competing risks model is employed for the analysis. The issue of separation of the outcomes (which precludes maximum likelihood estimation) is handled through the use of Bayesian inference with an appropriately chosen prior. We are interested in identifying important determinants of university outcomes and the associated model uncertainty is formally addressed through Bayesian model averaging. The methodology introduced for modelling university outcomes is applied to three selected degree programmes, which are particularly affected by dropout and late graduation.
\end{abstract}

\keywords{Bayesian model averaging; Competing risks; Outcomes separation; Proportional Odds model; University dropout}\endkeywords

\section{INTRODUCTION}

During the last decades, the  higher education system has seen substantial growth in Chile, evolving from around 165,000 students in the early 1980's to over 1 million enrolled in 2012 (see http://www.mineduc.cl/). Nowadays, the access to higher education is not restricted to an elite group. Among other reasons, this is due to a bigger role for education as a tool for social mobility, the opening of new institutions and a more accessible system of student loans and scholarships. However, currently, more than half of the students enrolled at Chilean higher education institutions do not complete their degree. This figure includes students expelled 
for academic or disciplinary reasons and those who voluntarily withdrew (dropout that is not instigated by the university but is also not necessarily the student's preference; \emph{e.g}. a student can be forced to drop out because of financial hardship). Another issue is the high frequency of late graduations, where obtaining the degree takes longer than the nominal duration of the programme. Chilean universities allow more flexibility than is usual in the Anglo-Saxon educational system, so students can repeat failed modules and/or have a reduced academic load in some semesters. Dropout and delays in graduation involve a waste of time and resources from the perspective of the students, their families, universities and the society.

There is a large literature devoted to university dropout. It includes conceptual models based on psychological, economic and sociological theories \citep[\emph{e.g.}][]{tinto1975,bean1980}. Here, instead, the focus is on empirical models. Previous research often considered the dropout as a dichotomous problem, neglecting the temporal component and focusing on whether or not a student has dropped out at a given time. Ignoring when the dropout occurs is a serious waste of information \citep{willettsinger1991}. Potential high risk periods will not be identified and no distinction between early and late dropout will be made. An alternative is to use (standard) survival models for the time to dropout \citep[as in][]{murtaughetat1999}, labelling graduated students as right censored observations. This is a major pitfall. Whilst students are enrolled at university, dropout is a possibility. However, dropout cannot occur after graduation, contradicting the idea of censoring. Instead, graduation must be considered as a competing event and incorporated into the survival model. 

We wish to identify determinants of the length of enrollment at university and its associated academic outcome for undergraduate students of the Pontificia Universidad Cat\'olica de Chile (PUC), which is one of the most prestigious universities in Chile (and the second best university in Latin America, according to QS Ranking 2013, see http://www.topuniversities.com/). Despite having one of the lowest dropout rates in the county (far below the national level), dropout is still an important issue for some degrees of the PUC. This goal of this analysis is to help university authorities to better understand the issue. Hopefully, it will also inspire policies mitigating late graduations and dropouts.

A competing risks model is proposed for the length of stay at university, where the possible events are: graduation, voluntary dropout and involuntary dropout. These are defined as the final academic situation recorded by the university at the end of 2011 (students that have not experienced any of these events by then are considered right-censored observations and censoring is assumed to be non-informative). In Chile, the academic year is structured in semesters (March-July and August-December). Survival times are defined as the length of enrollment at university and measured in semesters from admission (which means they are inherently discrete). It is an advantage of this approach that it deals jointly with graduations and dropouts. We aim to provide a better understanding of the problem for three selected programmes and to introduce a practically useful methodological framework.

The construction and the  main features of the PUC dataset are summarized in Section \ref{ChapPUCSectionDataset}, showing high levels of heterogeneity between programmes. This diversity is in terms of academic outcomes and the population in each degree programme. Section \ref{ChapPUCSectionCR} introduces a competing risks model for university outcomes, which can be estimated through a multinomial logistic regression. Bayesian inference is particularly helpful in this context where maximum likelihood inference is precluded. Section \ref{ChapPUCSectionImplementation} proposes a suitable prior structure, which is easy to elicit and introduces a Markov chain Monte Carlo (MCMC) algorithm that exploits a hierarchical representation of the multinomial logistic likelihood \citep[based on][]{holmesheld2006,polsonetal2013}. This section also proposes Bayesian model averaging to tackle model uncertainty in terms of the covariates used. The empirical results are summarized in Section \ref{ChapPUCSectionAnalysis}, focusing on some of the science programmes which are more affected by dropout and late graduations. Finally, Section \ref{ChapPUCSectionConcluding} concludes. R code is freely available at \newline \texttt{ http://www.warwick.ac.uk/go/msteel/steel\_homepage/software/university\_codes.zip} and is documented in the Supplementary Material, which also contains further descriptive analysis of the data and more computational details.

\section{THE PUC DATASET} \label{ChapPUCSectionDataset}

The PUC provided anonymized information about 34,543 undergraduate students enrolled during the period 2000-2011 via the ordinary admission process (based on high school marks and the results of a standardized university selection test, which is applied at a national level). 
Only the degree programmes that existed during the entire sample period are analyzed. In addition, we only consider students who: \emph{(i)} were enrolled for at least 1 semester (the dropout produced right after enrollment might have a different nature), \emph{(ii)} were enrolled in a single programme (students doing parallel degrees usually need more time to graduate and have less risk of dropout), \emph{(iii)} did not have validated previously passed modules from other degree programmes (which could reduce the time to graduation), \emph{(iv)} were alive by the end of 2011 (0.1\% of the students had died by then) and \emph{(v)} had full covariate information. Overall, 78.7\% of the students satisfied these criteria. The Supplementary Material breaks this number down by program. Throughout, we will only consider this subset of the data, pertaining to 27,185 students.

\begin{figure}[h!]
\begin{center}
\includegraphics[width=1\textwidth]{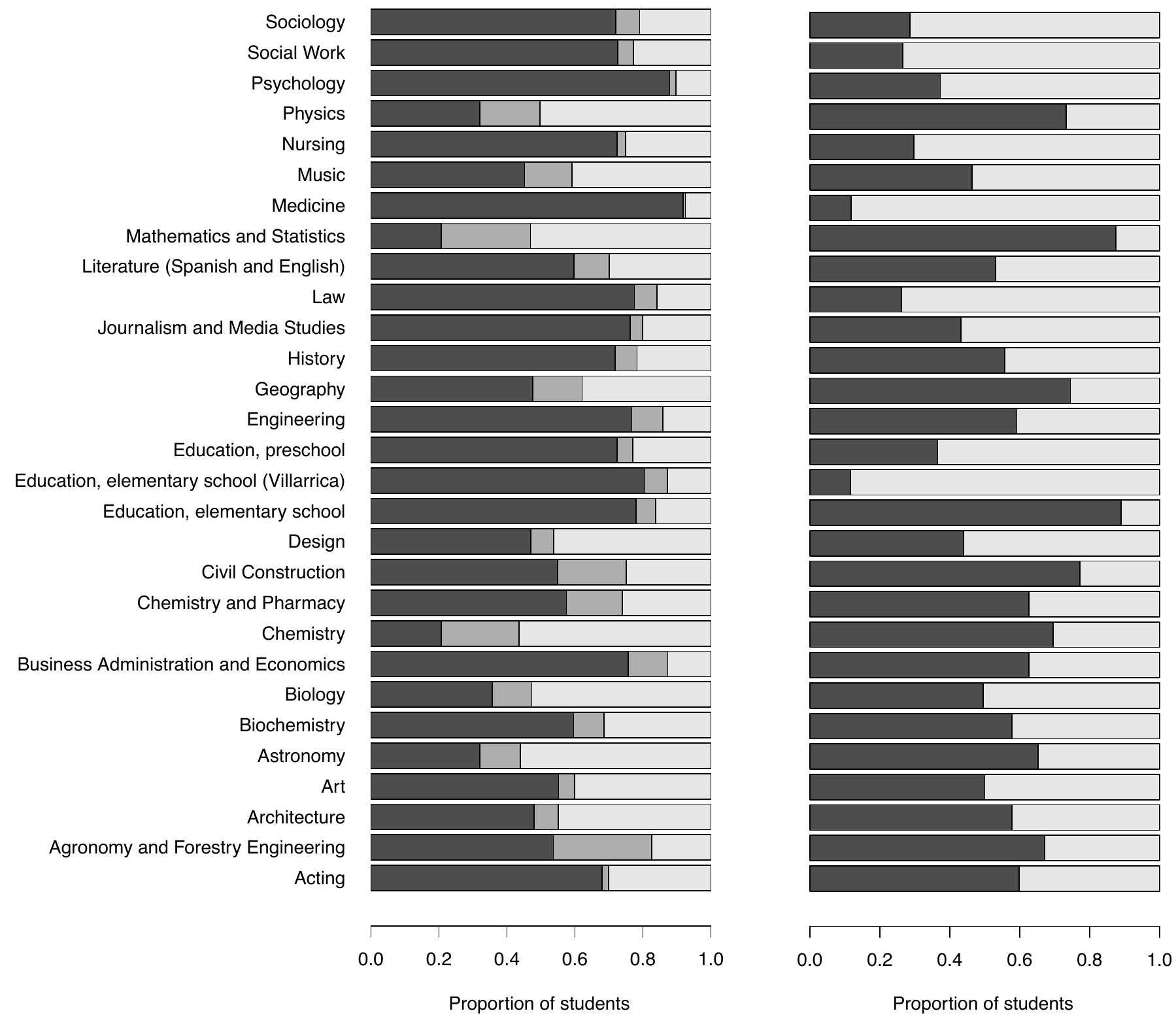}
\caption{\small Left: distribution of students according to final academic situation. From darkest to lightest, shaded areas represent the proportion of graduation, involuntary dropout and voluntary dropout, respectively. Right: distribution of graduated students according to timely graduation (within the nominal duration of the programme). The lighter area represents timely graduation.}
\label{graphPUCbarplots}
\end{center}
\end{figure}

By the end of 2011, 41.9\% of the students were still enrolled (right censored), 37.2\% had graduated, 6.6\% were expelled (involuntary dropout, mostly related to poor academic performances), 10.7\% withdrew (voluntary dropout), and 3.7\% abandoned the university without an official withdrawal. Following university policy, the latter group is classified as voluntary dropout. The high percentage of censoring mostly relates to students from later years of entry, who were not yet able to graduate by the end of 2011. From those who were not enrolled at the end of 2011, only 65\% had graduated (overall). The performance of former students is not homogenous across programmes (Figure \ref{graphPUCbarplots}). In terms of total dropout, Medicine (8.2\%) and Chemistry (79.4\%) have the lowest and highest rates, respectively. The highest rates of involuntary and voluntary dropout are for Agronomy and Forestry Engineering (28.9\%) and Chemistry (56.5\%), respectively. Dropouts are mostly observed during the first semesters of enrollment. In contrast, graduation times are concentrated on large values, typically above the official length of the programme (which varies between 8 and 14 semesters, with a typical value of 10 semesters). As shown in Figure \ref{graphPUCbarplots}, programmes also exhibit strong heterogeneity in terms of timely graduation, the proportion of which varies from 88\% (Medicine) to 11\% (Education Elementary School).

Demographic, socioeconomic and variables related to the admission process are recorded (see Table \ref{tableChapPUCCovariates}). For these covariates, substantial differences are observed between programmes (see Supplementary Material). In terms of demographic factors, some degrees have a very high percentage of female students (\emph{e.g.}~all education-related programmes) while {\it e.g.}~most of the Engineering students are male. The proportion of students who live outside the Metropolitan area is more stable across programmes (of course, a particularly high percentage is observed in the Education for Elementary School degree taught in the Villarrica campus, which is located in the south of Chile). Strong differences are also detected for the socioeconomic characterization of the students. Chilean schools are classified according to their funding system as public (fully funded by the government), subsidized private (the state covers part of the tuition fees) and private (no funding aid). This classification can be used as a proxy for the socioeconomic situation of the student (low, middle and upper class, respectively). The educational level of the parents is usually a good indicator of socioeconomic status as well. Some degrees have a very low percentage of students that graduated from public schools (\emph{e.g.} Business Administration and Economics) and others have a high percentage of students whose parents do not have a higher degree (\emph{e.g.}~Education for Elementary School in Villarrica). In addition, a few programmes have low rates of students with a scholarship or student loan (\emph{e.g.}~Business Administration and Economics). Finally, ``top'' programmes (\emph{e.g.}~Medicine, Engineering) only admit students with the highest selection scores. For instance, in 2011, the lowest selection score in Arts was 603.75 but Medicine did not enroll any students with a score below 787.75. In the same spirit, these highly selective programmes only enrolled students that applied to it as a first preference.

This substantial heterogeneity (in terms of outcomes and covariates) precludes meaningful modelling across programmes. Thus, the analysis will be done separately for each degree. 

\begin{table}[h!]
\caption{\small Available covariates (recorded at enrollment). Options for categorical variables in parentheses} \label{tableChapPUCCovariates} \centering
\begin{tabular}{p{14cm}}
  \hline
  \textbf{Demographic factors}\\
  \hline
  Sex (female, male)  \\
  Region of residence (Metropolitan area, others)   \\
  \hline
  \textbf{Socioeconomic factors} \\
  \hline
  Parents education (at least one with a technical or university degree, no degrees)\\
  High school type (private, subsidized private, public) \\
  Funding (scholarship and loan, loan only, scholarship only, none)  \\
  \hline
  \textbf{Admission-related factors} \\
  \hline
  Selection score   \\
  Application preference (first, others)  \\
  Gap between high school graduation and admission to PUC (1 year or more, none) \\
 \hline
\end{tabular}
\end{table}

\section{DISCRETE TIME COMPETING RISKS MODELS} \label{ChapPUCSectionCR}

Standard survival models only allow for a unique event of interest. Occurrences of alternative events are often recorded as censored observations. In the context of university outcomes, graduated students have been treated as censored observations when the event of interest is dropout \citep[as in][]{murtaughetat1999}. However, those students who graduated are obviously no longer at risk of dropout (from the same degree). Competing risks  models are more appropriate when several types of event can occur and there is a reason to believe they are the result of different mechanisms. These models simultaneously incorporate both the survival time and the type of event. Most of the previous literature focuses on continuous survival times \citep[\emph{e.g.}][]{crowder2001,pintilie2006}. Instead, in the context of university outcomes (where survival times are usually measured in numbers of academic terms), a discrete time approach is more appropriate. In a discrete-time competing risks setting, the variable of interest is $(R,T)$, where $R$ $\in \{1,\ldots,\mathcal{R}\}$ denotes the type of the observed event and $T \in\{1,2,\ldots\}$ is the survival time. Analogously to the single-event case, a model can be specified via the sub-distribution or sub-hazard functions, defined respectively as
\begin{equation}  F(r,t)  =  P(R=r,T \leq t) \,\,\mbox{and}\,\,
h(r,t) = \frac{P(R=r,T=t)}{P(T \geq t)} \label{eqChapPUCSubHazard}.
\end{equation}
The sub-distribution function $F(r,t)$ represents the proportion of individuals for which an event type $r$ has been observed by time $t$. The sub-hazard rate $h(r,t)$ is the conditional probability of observing an event of type $r$ during period $t$ given that no event (nor censoring) has happened before. The total hazard rate for all causes is defined as $h(t)=\sum_{r=1}^{\mathcal{R}} h(r,t)$. Like the Kaplan-Meier estimator in the discrete case, the maximum likelihood (non-parametric) estimator of $h(r,t)$ is the ratio between the number of events of type $r$ observed at time $t$ and the total number of individuals at risk at time $t$ \citep{crowder2001}. 

\begin{figure}
\begin{center}
\includegraphics[width=0.9\textwidth]{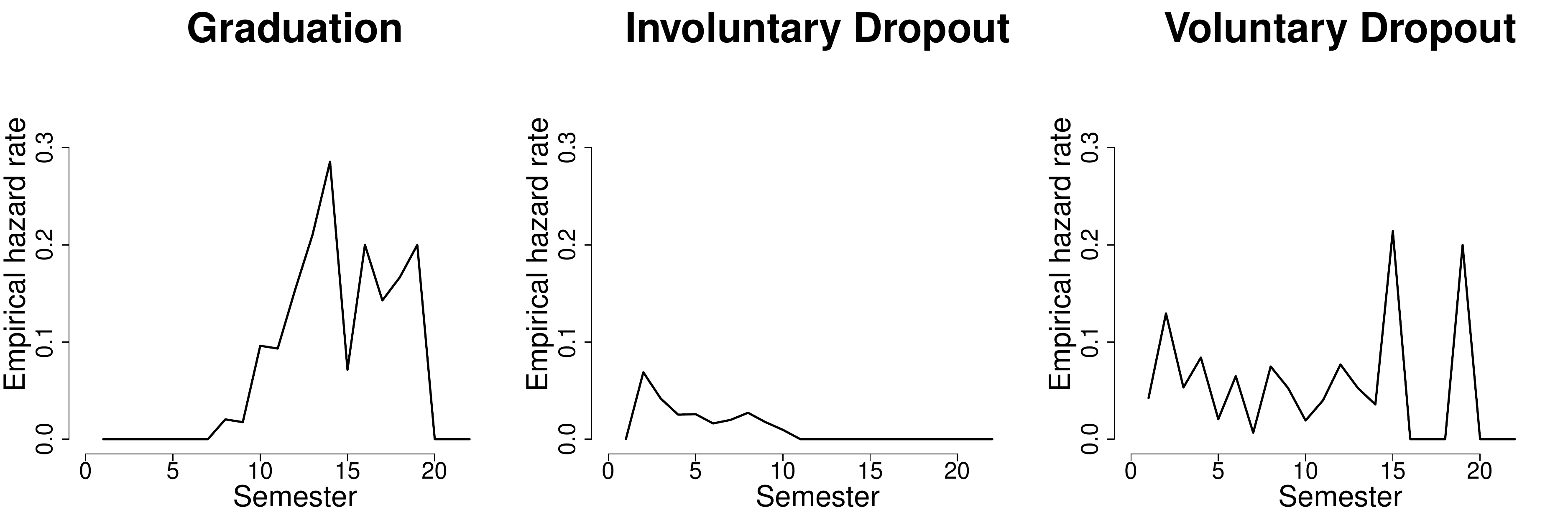}
\caption{\small Non-parametric estimation of cause-specific hazard rates for Chemistry students.}
\label{graphChapPUCHazardsChemistry}
\end{center}
\end{figure}

Sometimes, a simple (cause-specific) parametric model can be adopted. However, such 
models are not suitable for  the PUC dataset. For these data, the cause-specific hazard rates have a rather erratic behaviour over time. Figure \ref{graphChapPUCHazardsChemistry} illustrates this for Chemistry students. In particular, no graduations are observed during the first semesters of enrollment, inducing a zero graduation hazard at those times. Graduations only start about a year before the official duration of the programme (10 semesters). For this programme, the highest risk of being expelled from university is at the end of the second semester. In addition, during the first years of enrollment, the hazard of voluntary dropout has spikes located at the end of each academic year (even semesters). Therefore, more flexible models are required in order to accommodate these hazard paths.

\subsection{Proportional Odds model for competing risks data} \label{ChapPUCSectionModel}

\cite{cox1972} proposed a Proportional Odds (PO) model for discrete times and a single cause of failure. It is a discrete variation of the well-known Cox Proportional Hazard model, proposed in the same seminal paper. Let $x_i \in \mathbb{R}^k$ be a vector containing the value of $k$ covariates for individual $i$ while $\beta=(\beta_1,\ldots,\beta_k)' \in \mathbb{R}^k$ is a vector of regression parameters. The Cox PO model is given by \begin{equation} \label{eqChapPUCCoxPO} \log\left(\frac{h(t|\delta_{t},\beta;x_i)}{1-h(t|\delta_{t},\beta;x_i)}\right) = \log\left(\frac{h(t)}{1-h(t)}\right) + x_i'\beta \equiv \delta_{t}+x_i'\beta, \hspace{0.5cm} i=1,\ldots,n,\end{equation} where $\left\{\delta_{1},\delta_{2},\ldots\right\}$ respectively denote the baseline log-odds at times $\left\{1,2,\ldots\right\}$  and $t=1,\ldots,t_i$. The model in (\ref{eqChapPUCCoxPO}) can be estimated by means of a binary logistic regression. 
Define  $Y_{i {t}}$ as 1 if the event is observed at time {$t$} for individual $i$ and 0 otherwise. 
The likelihood related to (\ref{eqChapPUCCoxPO}) coincides with the likelihood corresponding to independent Bernoulli trials \citep{singerwillett1993}, where the contribution to the likelihood of individual $i$ (data collection for this individual stops if the event is observed or right censoring is recorded) is given by
\begin{equation} \label{eqBernoulli} L_i=P(Y_{it_{i}}=y_{it_{i}},\cdots,Y_{i1}=y_{i1}) = h(t_i)^{y_{it_{i}}} \prod_{s=1}^{t_i} [1-h(s)]^{1-y_{is}}. \end{equation} Equivalently, defining $c_i=0$ if the survival time is observed (\emph{i.e.} $Y_{it_{i}}=1,Y_{i(t_{i}-1)}=0,\cdots,Y_{i1}=0$) and $c_i=1$ if right censoring occurs (with $t_i$ as the terminal time), we can express the likelihood contribution as \begin{equation} L_i= \left[\frac{h(t_i)}{1-h(t_i)}\right]^{1-c_i} \prod_{s=1}^{t_i} [1-h(s)],\end{equation} where hazards are defined by (\ref{eqChapPUCCoxPO}) and the $\delta_{t}$'s are estimated by adding binary variables to the set of covariates. Now let $B=\left\{\beta_{(1)},\ldots,\beta_{(\mathcal{R})}\right\}$ be a collection of cause-specific regression parameters (each of them in $\mathbb{R}^k$) and define $\delta=\left\{\delta_{1 1},\ldots,\delta_{\mathcal{R} 1},\delta_{1 2},\ldots,\delta_{\mathcal{R} 2},\ldots\right\}$. The model in (\ref{eqChapPUCCoxPO}) can then be extended in order to accommodate $\mathcal{R}$ possible events via the following multinomial logistic regression model \begin{eqnarray} \label{eqChapPUCMultinomial1} \log\left(\frac{h(r,t|\delta,B; x_i)}{h(0,t|\delta,B;x_i)}\right) & = &  \delta_{r t}+x_i'\beta_{(r)},\hspace{0.5cm}r=1,\ldots,\mathcal{R}; { t=1,\ldots,t_i}; i=1,\ldots,n, \\  \mbox{where\hspace{0.2cm}} h(0,t|\delta,B; x_i) & = & 1 -\sum_{r=1}^{\mathcal{R}} h(r,t|\delta,B; x_i) \end{eqnarray} is the hazard of no event being observed at time $t$. The latter is equivalent to \begin{equation} \label{eqChapPUCMultinomial2} h(r,t|\delta,B;x_i) = \frac{\,e^{\delta_{r t}+ x_i'\beta_{(r)}}}{1+\sum_{s=1}^{\mathcal{R}} \,e^{\delta_{s t}+ x_i'\beta_{(s)}}}.\end{equation}

This notation implies that the same predictors are used for each cause-specific component (but this is easily generalised). In (\ref{eqChapPUCMultinomial1}), covariates influence both the marginal probability of the event $P(R=r)$ and the rate at which the event occurs. Positive values of the cause-specific coefficients indicate that (at any time point) the hazard of the corresponding event increases with the associated covariate values and the effect of covariates on log odds is constant over time. For university outcomes, (\ref{eqChapPUCMultinomial1}) has been used by \cite{scottkennedy2005}, \cite{ariasortiseanddehon2011} and \cite{clericietall2013}, among others. Nonetheless, its use has some drawbacks. Firstly, it involves a large number of parameters (if $\mathcal{T}$ is the largest recorded time, there are $\mathcal{R} \times \mathcal{T}$ different $\delta_{rt}$'s). \cite{scottkennedy2005} overcome this by assigning a unique indicator $\delta_{rt_0}$ to the period $[t_0,\infty)$ (for fixed $t_0$). The choice of $t_0$ is rather arbitrary but it is reasonable to choose $t_0$ such that most individuals already experienced one of the events (or censoring) by time $t_0$. Secondly, maximum likelihood inference for the multinomial logistic regression is precluded when the outcomes are (quasi) completely separated with respect to the predictors, \emph{i.e.}~some outcomes are not (or rarely) observed for particular covariate configurations \citep{albertanderson1984}. In other words, the predictors can (almost) perfectly predict the outcomes. In (\ref{eqChapPUCMultinomial1}), these predictors include binary variables representing the period indicators $\delta_{rt}$'s. Therefore, (quasi) complete separation occurs if the event types are (almost) entirely defined by the survival times. This is a major issue in the context of university outcomes. For example, no graduations can be observed during the second semester of enrollment. Therefore, the likelihood function will be maximized when the cause-specific hazard related to graduations (defined in (\ref{eqChapPUCMultinomial2})) is equal to zero at time $t=2$. Thus, the ``best'' value of the corresponding period-indicator is $-\infty$.

\cite{singerwillett2003} use polynomial baseline odds to overcome the separation issue. This option is less flexible than (\ref{eqChapPUCMultinomial1}), and its use is only attractive when a low-degree polynomial can adequately  represent the baseline hazard odds. This is not the case for the PUC dataset, where cause-specific hazard rates have a rather complicated behaviour (see Figure \ref{graphChapPUCHazardsChemistry}) and not even high-order polynomials would provide a good fit.

Here, the model in (\ref{eqChapPUCMultinomial1}) is adopted for the analysis of the PUC dataset, using Bayesian methods to handle separation. We define the last period as $[t_0,\infty)$ \citep[for fixed $t_0$, as in][]{scottkennedy2005}, and period-indicators for time $t=1$  are defined as cause-specific intercepts. 

\section{BAYESIAN PO COMPETING RISKS REGRESSION} \label{ChapPUCSectionImplementation}

\subsection{Prior specification}

\begin{figure}
\begin{center}
\includegraphics[width=0.9\textwidth]{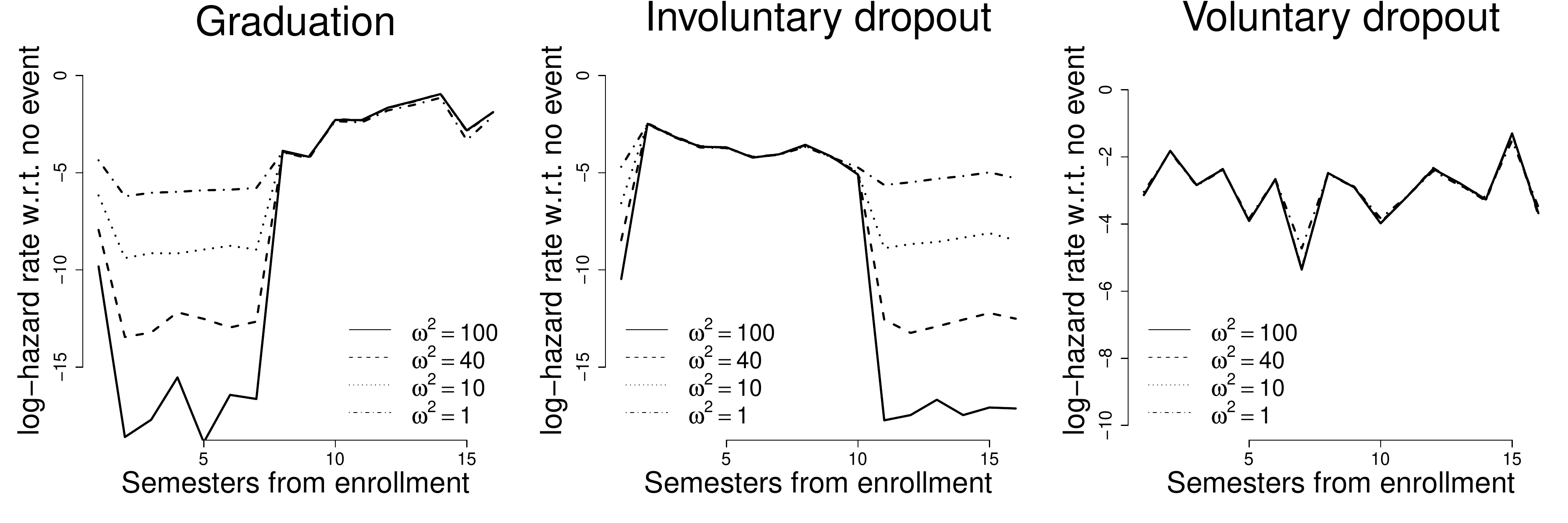}
\caption{\small For Chemistry students: posterior median trajectory of the log-hazard ratio for each competing event with respect to no event being observed, using the model in (\ref{eqChapPUCMultinomial1}) under $\delta_r \sim$ Cauchy$_{{t_0}}$($0\iota_{t_0},\omega^2 I_{{t_0}}$).} 
\label{graphChapPUCTimePeriodChemistry}
\end{center}
\end{figure}

An alternative solution to the separation issue lies in the Bayesian paradigm, allowing the extraction of information from the data via an appropriate prior distribution for the period-indicators $\delta_{rt}$ \citep{gelmanetal2008}. The Jeffreys prior can be used for this purpose \citep{firth1993}. This is attractive when reliable prior information is absent. In a binary logistic case, the Jeffreys prior is proper and its marginals are symmetric with respect to the origin \citep{ibrahimlaud1991,poirier1994}. These properties have no easy generalization for the multinomial case, where an expression for the Jeffreys prior is very complicated \citep{poirier1994}. Instead, \cite{gelmanetal2008} suggested weakly informative independent Cauchy priors for a re-scaled version of the regression coefficients. When the outcome is binary, these Cauchy (and any Student $t$) priors are symmetric like the Jeffreys prior but produce fatter tails \citep{chenetal2008}. The prior in \cite{gelmanetal2008} assumes that the regression coefficients fall within a restricted range. For the model in (\ref{eqChapPUCMultinomial1}), it penalizes large differences between the $\delta_{rt}$'s associated with the same event. Such a prior is convenient if the separation of the outcomes relates to a small sample size (and increasing the sample size will eventually eliminate this issue). This is not the case for the PUC dataset, or other typical data on university outcomes,  where the separation arises from structural restrictions (\emph{e.g.}~it is not possible to graduate during the first periods of enrollment). Hence, large differences are expected for the $\delta_{rt}$'s associated with the same event. In particular,  $\delta_{rt}$ should have a large negative value in those periods where event $r$ is very unlikely to be observed (inducing a nearly zero cause-specific hazard rate). Defining $\delta_r=(\delta_{r1},\ldots,\delta_{r t_0})'$, we suggest the prior
\begin{equation} \label{eqChapPUCCauchyprior} \delta_r \sim \mbox{Cauchy}_{t_0}(0 \iota_{t_0},\omega^2 I_{{t_0}}), \hspace{0.5cm} r=1,\ldots,\mathcal{R}\end{equation} where $I_{{t_0}}$ denotes the identity matrix of dimension ${t_0}$ and $\iota_{t_0}$ is a vector of $t_0$ ones. Equivalently, \begin{equation} \label{eqChapPUCCauchyprior2} \pi(\delta_{r}|\Lambda_r=\lambda_{r}) \sim \mbox{Normal}_{{t_0}}(0 \iota_{t_0}, \lambda_r^{-1} \omega^2  I_{{t_0}}),\hspace{0.5cm} \Lambda_r \sim \mbox{Gamma}(1/2,1/2),\hspace{0.5cm} r=1,\ldots,\mathcal{R}. \end{equation}
This prior assigns non-negligible probability to large negative (and positive) values of the $\delta_{rt}$'s. Of course, an informative prior could also be used, but this would require non-trivial prior elicitation and it is not entirely clear a priori which $\delta_{rt}$'s are affected by the separation issue. Focusing on Chemistry students and using different values of $\omega^2$ for the prior in (\ref{eqChapPUCCauchyprior}), Figure \ref{graphChapPUCTimePeriodChemistry} shows the induced trajectory for the posterior median of the log-hazard ratio for each event type with respect to no event being observed. For simplicity, covariates are excluded for this comparison. Choosing a value of $\omega^2$ is not critical for those periods where the separation is not a problem (as the data is more informative). In contrast, $\omega^2$ has a large effect in those semesters where the separation occurs. Tight priors \citep[as the ones in][]{gelmanetal2008} are too conservative and produce non-intuitive results. Hence, large values of $\omega^2$ seem more appropriate. How large is arbitrary but, after a certain threshold, its not too relevant in the hazard ratio scale (as the hazard ratio will be practically zero). For the analysis of the PUC dataset, $\omega^2=100$ is adopted.

The Bayesian model is completed using independent $g$-priors \citep{zellner1986} for the cause-specific regression  coefficients, \emph{i.e.} \begin{equation} \label{eqChapPUCgprior} \beta_{(r)} \sim \mbox{Normal}_{k}(0 \iota_k,g_{r}(X'X)^{-1}), \hspace{1cm} r=1,\ldots,\mathcal{R}, \end{equation} where $X=(x_1,\dots,x_n)'$.  This is a popular choice in Bayesian model selection and averaging under uncertainty regarding the inclusion of covariates \citep[\emph{e.g.}][]{fernandezetal2001}. This prior is invariant to scale transformations of the covariates. The particular choice of fixed values for  $\left\{g_{1},\ldots,g_{\mathcal{R}}\right\}$ can fundamentally affect the posterior inference \citep{liangetal2008,leysteel2009}. For a binary logistic regression, \cite{hansonetal2014}  elicit $g_r$ using averaged prior information (across different covariates configurations). Alternatively, a hyper-prior can be assigned to each $g_r$, inducing a hierarchical prior structure \citep{liangetal2008}. Several choices for this hyper-prior are examined in \cite{leysteel2012}. Based on theoretical properties and a simulation study (in a linear regression setting) they recommend a benchmark Beta prior for which \begin{equation} \label{eqChapPUCBenchmarkBeta} \frac{g_r}{1+g_r}  \sim  \mbox{Beta}(b_1,b_2) \hspace{0.5cm} \mbox{or equivalently} \hspace{0.5cm}
\pi(g_r) = \frac{\Gamma(b_1+b_2)}{\Gamma(b_1)\Gamma(b_2)} g_r^{b_1-1} (1+g_r)^{-(b_1+b_2)},\end{equation} where $b_1=0.01\max\{n,k^2\}$ and $b_2=0.01$. The prior in (\ref{eqChapPUCgprior}) and  (\ref{eqChapPUCBenchmarkBeta}) is adopted for the regression coefficients throughout the analysis of the PUC dataset.

\subsection{Markov chain Monte Carlo implementation}

Fitting a multinomial (or binary) logistic regression is not straightforward. There is no conjugate prior and sampling from the posterior distribution is cumbersome \citep{holmesheld2006}. The Bayesian literature normally opts for alternative representations of the multinomial logistic likelihood. For instance, \cite{forster2010} exploits the relationship between a multinomial logistic regression and a Poisson generalized linear model. Following \cite{albertandchib1993}, \cite{holmesheld2006} adopt a hierarchical structure where the logistic link is represented as a scale mixture of normals. Alternatively, \cite{fruhwirth2010} approximated the logistic link via a finite mixture of normal distributions. In the present paper, the hierarchical structure proposed in \cite{polsonetal2013} is adapted for our model. For a binary logistic model with observations  $\left\{y_{it}: i=1,\ldots,n, t=1,\ldots,t_i\right\}$ ($y_{it}=1$ if the event is observed at time $t$ for subject $i$, $y_{it}=0$ otherwise), the key result in \cite{polsonetal2013} is that \begin{equation} \label{eqChapPUCPolson}\frac{ [\,e^{z'_i \beta^*}]^{y_{it}}}{\,e^{z'_i \beta^*}+1}  \propto \,e^{\kappa_{it} z'_i \beta^*} \int_0^{\infty} \exp\{-\eta_{it}(z'_i \beta^*)^2/2\} f_{PG}(\eta_{it}|1,0) \,d \eta_{it}, \end{equation} where $z_i$ is a vector of covariates associated with individual $i$, $\beta^*$ is a vector of regression coefficients, $\kappa_{it}=y_{it}-1/2$ and $f_{PG}(\cdot|1,0)$ denotes a Polya-Gamma density with parameters 1 and 0. In terms of the model in (\ref{eqChapPUCCoxPO}), $z_i$ includes $x_i$ and the binary indicators linked to the $\delta_t$'s. Thus, $\beta^*=(\delta_1,\ldots,\delta_{t_0},\beta')'$.

The result in (\ref{eqChapPUCPolson}) can be used to construct a Gibbs sampling scheme for the multinomial logistic model along the lines of \cite{holmesheld2006}. Now let $0,1,\ldots,\mathcal{R}$ be the possible values for observations $y_{it}$ associated with regression coefficients $\beta^*_{(1)},\ldots,\beta^*_{(\mathcal{R})}$. Given $\beta^*_{(1)},\ldots,\beta^*_{(r-1)},\beta^*_{(r+1)},\ldots,\beta^*_{(\mathcal{R})}$, the ``conditional'' likelihood function for $\beta^*_{(r)}$ is proportional to
\begin{equation} \label{eqChapPUCHolmes} \prod_{i=1}^n \prod_{t=1}^{t_i} \frac{\left[\exp\{z_i'\beta^*_{(r)}-C_{ir}\}\right]^{I(y_{it}=r)}}{1+\exp\{z_i'\beta^*_{(r)}-C_{ir}\}}, \mbox{\hspace{0.2cm}where \hspace{0.2cm}} C_{ir}  = \log\left(1+\sum_{r^* \neq r} \exp\{z_i'\beta^*_{(r^*)}\}\right). \end{equation} Assume $\beta^*_{(r)} \sim \mbox{Normal}_{{t_0}+k}\left(\mu_{r},\Sigma_{r}\right)$, $r=1,\ldots,\mathcal{R}$ and define $B^*=\left\{\beta^*_{(1)},\ldots,\beta^*_{(\mathcal{R})}\right\}$. Using (\ref{eqChapPUCPolson}) and (\ref{eqChapPUCHolmes}), a Gibbs sampler for the multinomial logistic model is defined through the following full conditionals for $r=1,\ldots,\mathcal{R}$\begin{eqnarray} && \beta^*_{(r)}|\eta_r, \beta^*_{(1)},\ldots,\beta^*_{(r-1)},\beta^*_{(r+1)},\ldots,\beta^*_{(\mathcal{R})}, y_{11}\dots,y_{nt_n}  \sim  \mbox{Normal}_{{t_0}+k}(m_{r},V_{r}),  \\
&&\eta_{itr}|B^*  \sim  \mbox{PG}(1,z'_i \beta^*_{(r)}-C_{ir}), \hspace{0.2cm} t=1,\dots,t_i, i=1,\ldots,n, \end{eqnarray} defining  $Z=(z_1 \otimes \iota_{t_1}',\ldots, z_n\otimes \iota_{t_n}')'$, $\eta_r=(\eta_{11r},\ldots,\eta_{nt_nr})'$, $D_{r}=\mbox{diag}\{\eta_r\}$, $V_{r}=(Z'D_{r} Z + \Sigma_{r}^{-1})^{-1}$, $m_{r}=V_{r}(Z'\kappa_{r}+\Sigma_{r}^{-1}\mu_{r})$, $\kappa_{r}=(\kappa_{11r},\ldots,\kappa_{nt_nr})'$ and $\kappa_{itr}=$ I$_{\{y_{it}=r\}}-1/2+\eta_{itr}C_{ir}$ (where I$_A=1$ if $A$ is true, $0$ otherwise). The previous algorithm applies to (\ref{eqChapPUCMultinomial1}) using $\beta^*_{(r)}=(\delta'_{r},\beta'_{(r)})'$ and defining $z_i$ in terms of binary variables related to the $\delta_{rt}$'s and the covariates $x_i$.
Extra steps are required to accommodate the adopted prior, which is a product of independent multivariate Cauchy and  hyper-$g$ prior components. Both components can be represented as a scale mixture of normal distributions (see (\ref{eqChapPUCCauchyprior2}) and (\ref{eqChapPUCgprior})). Hence, conditional on $\Lambda_1,\ldots,\Lambda_{\mathcal{R}},g_1,\ldots,g_{\mathcal{R}}$, the sampler above applies. In addition, at each iteration, $\Lambda_r$'s and $g_r$'s are updated using the full conditionals.
\begin{eqnarray} \Lambda_r|\delta_r & \sim & \mbox{Gamma}\left(\frac{{t_0}+1}{2},\frac{\delta'_r \delta_r}{2\omega^2}\right), \hspace{0.5cm} r=1,\ldots,\mathcal{R},  \\
g_r|\beta_r & \sim & g_r^{-k/2} \exp\left\{-\frac{\beta'_r X'X \beta_r}{2g_r} \right\} \pi(g_r), \hspace{0.5cm} r=1,\ldots,\mathcal{R}. \label{eqChapPUCfullcondg}\end{eqnarray} An adaptive Metropolis-Hastings step \citep[see Section 3 in][]{robertsrosenthal2009} is implemented for (\ref{eqChapPUCfullcondg}).

\subsection{Bayesian variable selection and model averaging}

A key aspect of the analysis is to select the relevant covariates to be included in the model. Often, a unique model is chosen via some model comparison criteria. The Deviance Information Criteria (DIC) of \cite{spiegelhalterelat2002} is computed. Low DIC suggests a better model. We also consider the Pseudo Marginal Likelihood (PsML) predictive criterion, proposed in \cite{geisserandeddy1979}. Higher values of PsML indicate a better predictive performance. The Supplementary Material (Section D) provides more details.

In a Bayesian setting, a natural way to deal with model uncertainty is through posterior model probabilities. Denote by $k^*$ the number of available covariates ($k^*$ might differ from the number of regression coefficients because categorical covariates may have more than two levels). Let $M_1,\ldots,M_\mathcal{M}$ be the set of all $\mathcal{M}=2^{k^*}$ competing models (if a categorical covariate is included, all its levels are incorporated). Given observed times $T_{obs}$ and event types $R_{obs}$, posterior probabilities for these models are defined via Bayes theorem as \begin{equation} \label{eqChapPUCPostProb} \pi(M_m|T_{obs},R_{obs}) = \frac{L(T_{obs},R_{obs}|M_m) \pi(M_m)}{\sum_{m^*=1}^{\mathcal{M}} L(T_{obs},R_{obs}|M_{m^*}) \pi(M_{m^*})}, \mbox{\hspace{0.1cm} with \hspace{0.1cm}} \sum_{m=1}^{\mathcal{M}} \pi(M_m)=1, \end{equation} where $\pi(M_1),\ldots,\pi(M_{\mathcal{M}})$ represent the prior on model space and $L(T_{obs},R_{obs}|M_m)$ is the marginal likelihood for model $m$ (computed as in Section C of the the Supplementary Material). A uniform prior on the model space is defined as \begin{equation} \label{eqChapPUCunifprior1} \pi(M_m) = \frac{1}{\mathcal{M}}, \hspace{0.5cm} m=1,\ldots, \mathcal{M}. \end{equation} Alternatively, a prior for the model space can be specified through the covariate-inclusion indicators $\gamma_j$, which take the value 1 if covariate $j$ is included and 0 otherwise, $j=1,\ldots,k^*$. Independent Bernoulli($\theta$) priors are assigned to the $\gamma_j$'s. For $\theta=1/2$, the induced prior coincides with the uniform prior in (\ref{eqChapPUCunifprior1}). As discussed in \cite{leysteel2009}, assigning a hyper-prior for $\theta$ provides more flexibility and reduces the influence of prior assumptions on posterior inference. A Beta($a_1,a_2$) prior for $\theta$ leads to the so-called Binomial-Beta prior on the number of included covariates $W=\sum_{j=1}^{k^*} \gamma_j$. If $a_1=a_2=1$ (uniform prior for $\theta$), the latter induces a uniform prior for $W$, \emph{i.e.} \begin{equation} \label{eqChapPUCunifprior2} \pi(W=w)=\frac{1}{k^*+1}, \hspace{0.5cm} w=0,\ldots,k^*. \end{equation}

A formal Bayesian response to inference under model uncertainty is Bayesian Model Averaging (BMA), which averages over all possible models with the posterior model probabilities, instead of selecting a single model. Surveys can be found in \cite{hoetingetal1999} and \cite{chipmanetal2001}. Let $\Delta$ be a quantity of interest (\emph{e.g.}~a covariate effect). Using BMA, the posterior distribution of $\Delta$ is given by \begin{equation}\label{eqChapPUCBMA} P(\Delta|T_{obs},R_{obs}) = \sum_{m=1}^{\mathcal{M}} P_m(\Delta|T_{obs},R_{obs}) \pi(M_m|T_{obs},R_{obs}),\end{equation} where $P_m(\Delta|T_{obs},R_{obs})$ denotes the posterior distribution of $\Delta$ for model $M_m$.  BMA has been shown to lead to better predictive performance than choosing a single model \citep{rafteryetal1997,fernandezetal2001}.

\section{EMPIRICAL RESULTS FOR THE PUC DATA}  \label{ChapPUCSectionAnalysis}

The PUC dataset is analyzed through the model in (\ref{eqChapPUCMultinomial1}) using the prior and the algorithm described in Section \ref{ChapPUCSectionImplementation}. As indicated in Section \ref{ChapPUCSectionDataset}, the analysis is carried out independently for each programme, focusing on some of the science programmes for which the rates of dropout and/or late graduations are normally higher. In particular, we consider Chemistry (379 students), Mathematics and Statistics (598 students) and Physics (237 students). For all programmes, 8 covariates are available (see Table \ref{tableChapPUCCovariates}), inducing $2^8=256$ possible models (using the same covariates for each cause-specific hazard). Selection scores cannot be directly compared across admission years (the test varies from year to year). Hence, the selection score is replaced by an indicator of being in the top 10\% of the enrolled students (for each programme and admission year). The following regression coefficients are defined for each cause (the subscript $r$ is omitted for ease of notation): $\beta_1$ (sex: female), $\beta_2$ (region: metropolitan area), $\beta_3$ (parents' education: with degree), $\beta_4$ (high school: private), $\beta_5$ (high school: subsidized private), $\beta_6$ (funding: scholarship only), $\beta_7$ (funding: scholarship and loan), $\beta_8$ (funding: loan only), $\beta_9$ (selection score: top 10\%), $\beta_{10}$ (application preference: first) and $\beta_{11}$ (gap after high school graduation: yes). All models contain an intercept and ${t_0}-1=15$ period indicators. For all models, the total number of MCMC iterations is 200,000 and results are presented on the basis of 1,000 draws (after a burn-in of 50\% of the initial iterations and thinning). Trace plots and the usual convergence criteria strongly suggest  good mixing and  convergence of the chains (not reported).

\begin{figure}[h!]
\begin{center}
\includegraphics[width=0.9\textwidth]{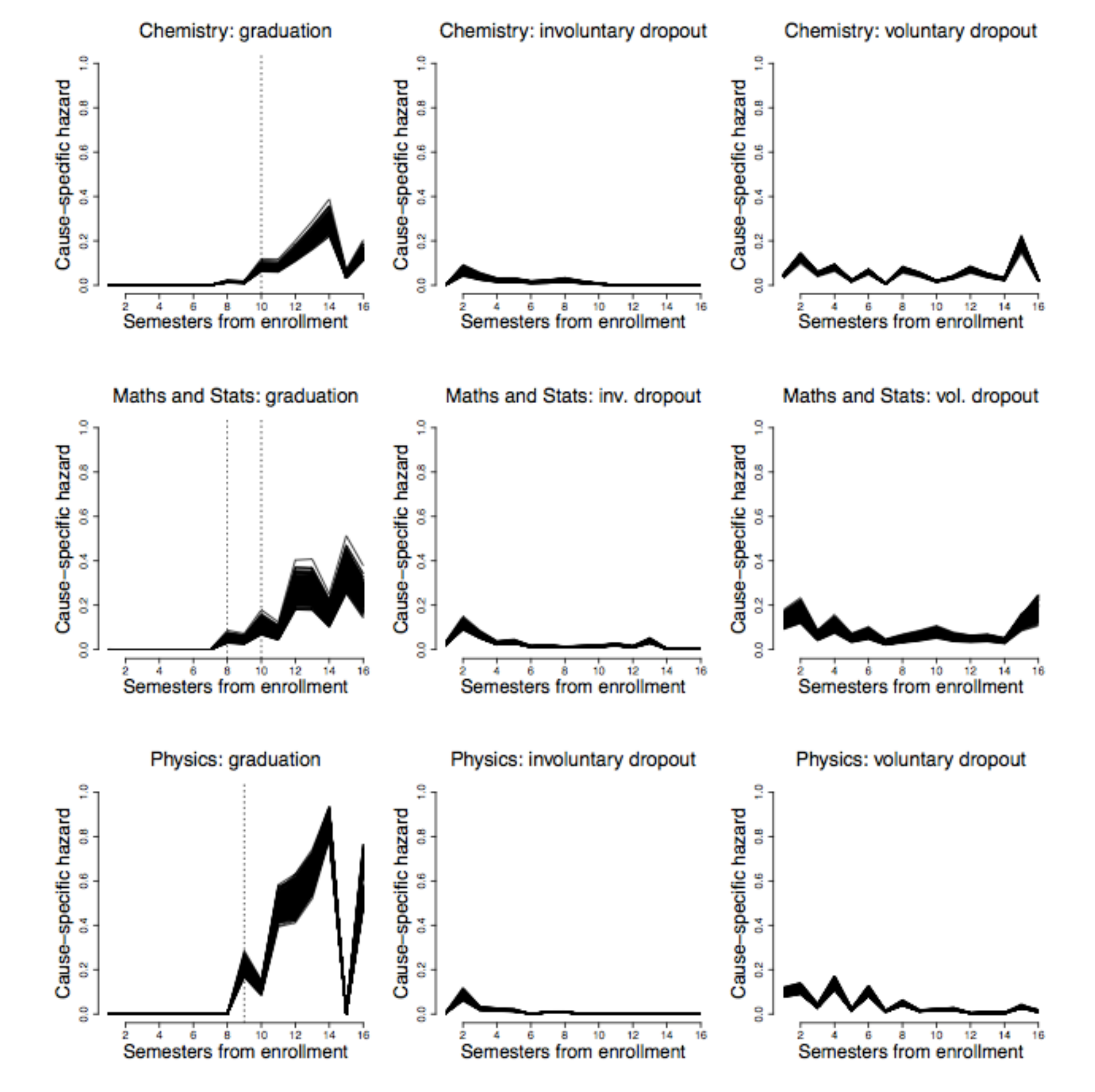}
\caption{\small Posterior medians of baseline cause-specific hazards (defined in terms of the $\delta_{rt}$'s) across the 256 possible models. For graduation hazards, dashed vertical lines are located at the official duration of the programme (in Mathematics and Statistics students in Statistics can take two additional semesters to get a professional degree).} \label{graphChapPUCHazards}
\end{center}
\end{figure}

Figure \ref{graphChapPUCHazards} displays the trajectory of the cause-specific hazard rates for all possible 256 models, corresponding to the reference case (where $x_i=0 \iota_k$). Differences between these estimations are mostly related to changes in the intercept, which is obviously affected by the removal or addition of covariates. The first row of panels in Figure \ref{graphChapPUCHazards} roughly recovers the same patterns as in Figure \ref{graphChapPUCHazardsChemistry}, suggesting that these estimates are dominated by the data and not by the prior. Some similarities appear between these programmes. For example, the highest risk of involuntary dropout is observed by the end of the second semester from enrollment. This may relate to a bad performance during the first year of studies. In addition, during the 4 first years of enrollment, the hazard rate associated to voluntary dropouts has spikes located at even semesters. Again, this result is intuitive. Withdrawing at the end of the academic year allows students to re-enroll in a different programme without having a gap in their academic careers. In terms of graduations, mild spikes are located at the official duration of the programmes. Nonetheless, for these programmes, the highest hazards of graduation occur about 4 semesters after the official duration. The spikes at the last period are due to a cumulative effect (as $\delta_{r t_0}$ represents the period $[t_0,\infty)$).

\begin{figure}[h!]
\begin{center}
\subfigure{\includegraphics[width=0.95\textwidth]{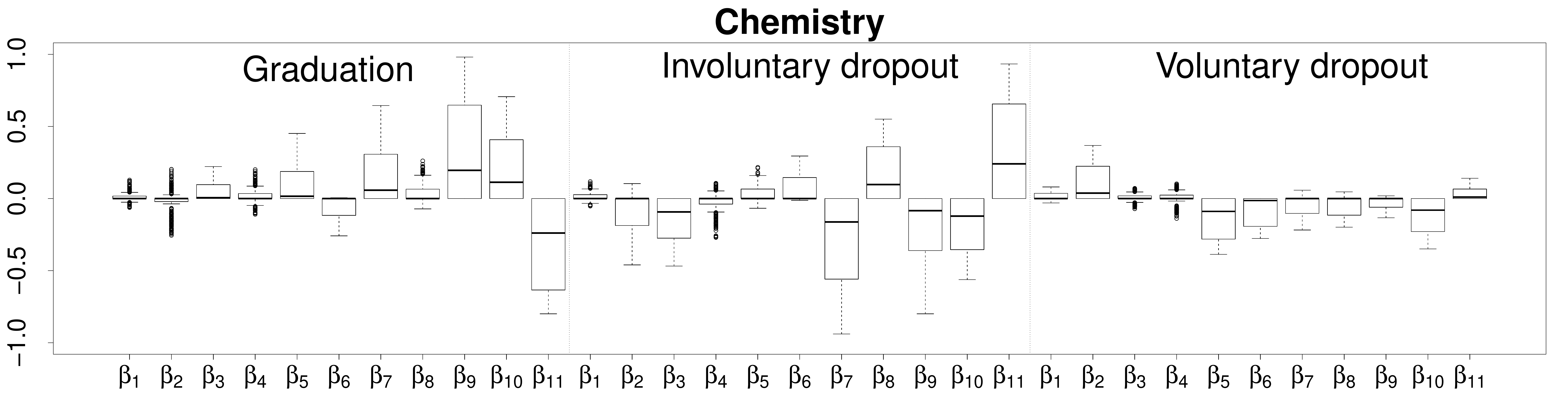}}\\
\subfigure{\includegraphics[width=0.95\textwidth]{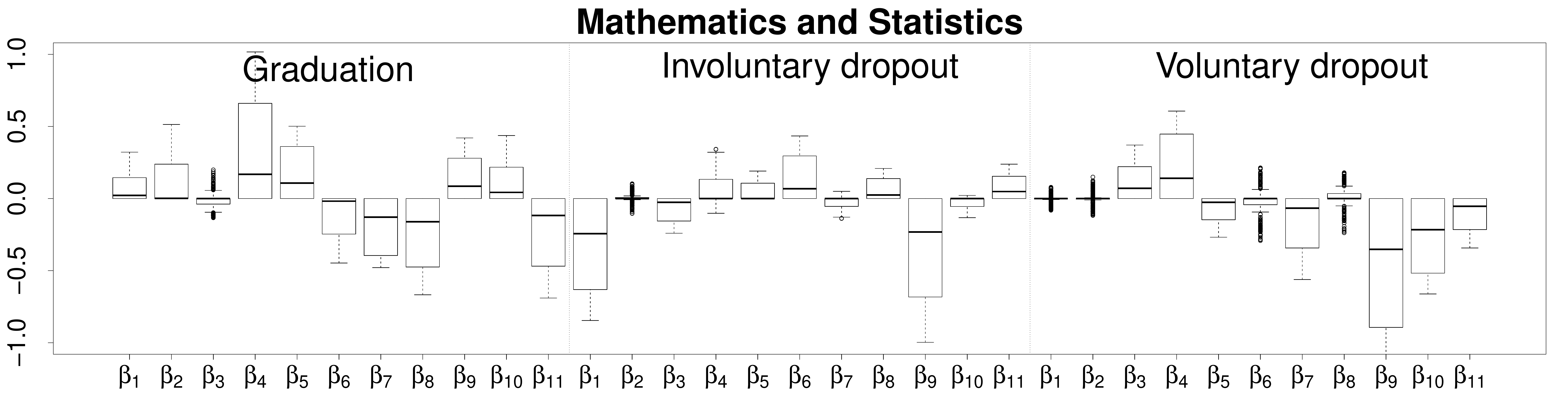}}\\
\subfigure{\includegraphics[width=0.95\textwidth]{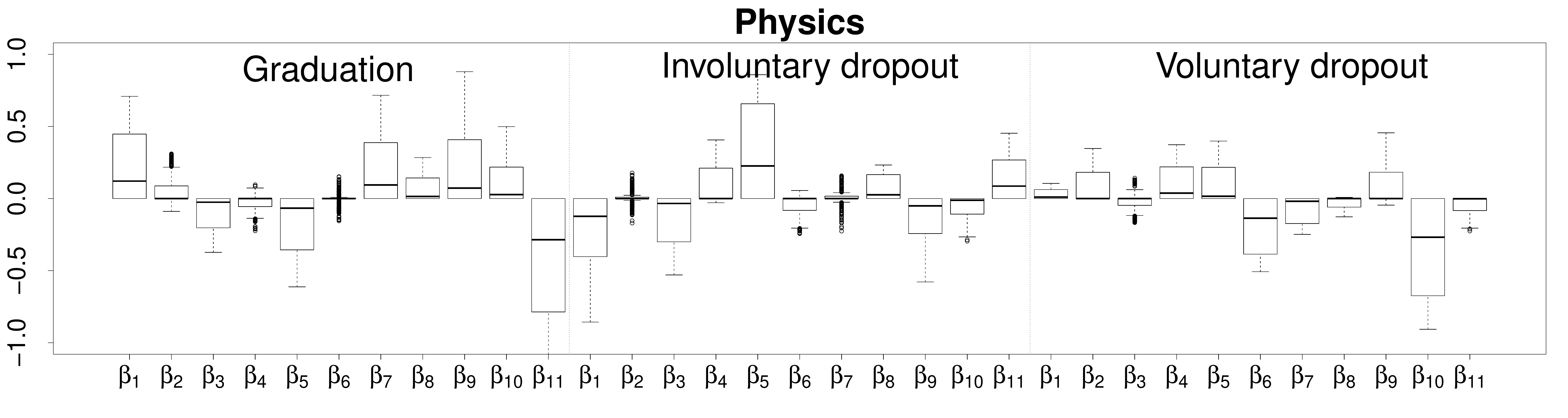}}
\caption{\small Boxplot of estimated posterior medians of covariate effects across the 256 possible models. When a covariate is not included in the model, the corresponding posterior median is zero.} \label{graphChapPUCEffects}
\end{center}
\end{figure}

Figure \ref{graphChapPUCEffects} summarizes marginal posterior inference under all possible 256 models. Across all models, the median effects normally retain the same sign (within the same programme). Only covariates with smaller effects display estimates with opposite signs (\emph{e.g.}~the coefficient related to sex, $\beta_1$, for Chemistry students). Nonetheless, the actual effect values do not coincide across different models. In general, students who applied as a first preference to these degrees graduated more and faster (see estimations of $\beta_{10}$). These students also exhibit a lower rate of voluntary dropout, which might be linked to a higher motivation. Whether or not the student had a gap between high school graduation and university admission also has a strong influence on the academic outcomes for these programmes. These gaps can, for example, correspond to periods in which the student was preparing for the admission test (after a low score in a previous year) or enrolled in a different programme. Overall, this gap induces less and slower graduations for these programmes. In addition, in each semester, students with a gap before university enrollment have a higher risk of being expelled from these degrees. The effects of the covariates are not homogeneous across the programmes. Whereas the effect of the student's sex ($\beta_1$) is almost negligible in Chemistry, female students in Mathematics and Statistics and in Physics present a higher hazard of graduation and lower risk of being expelled in all semesters.

\begin{table}[h!]
\caption{\small Top 3 models in terms of DIC and PsML (ticks indicate covariate inclusion).}\label{tableChapPUCDICPsML} \centering \small \tabcolsep=0.11cm
\begin{tabular}[c]{lccccccccc}
  \hline
  Programme & DIC & Sex	& Region & Parents & School & Funding & Top 10\% & Pref. & Gap \\
  \hline
  \multirow{3}{*}{Chemistry} & 1915.23 &  & $\checkmark$ &  &  &  & $\checkmark$ & $\checkmark$ & $\checkmark$ \\
                             & 1915.54 &  &  &  &  &  & $\checkmark$ & $\checkmark$ & $\checkmark$ \\
                             & 1915.64 &  &  &  &  &  &  & $\checkmark$ & $\checkmark$ \\
  \hline
  Mathematics                & 3117.89 &$\checkmark$	&	&	&$\checkmark$	&	&$\checkmark$	&$\checkmark$	&$\checkmark$  \\
  and                        & 3119.95 &$\checkmark$	&$\checkmark$	&	&$\checkmark$	&	&$\checkmark$	&$\checkmark$	&$\checkmark$  \\
  Statistics                 & 3120.06 &$\checkmark$	&	&$\checkmark$	&$\checkmark$	&	&$\checkmark$	&$\checkmark$	&$\checkmark$  \\
  \hline
  \multirow{3}{*}{Physics}   & 1091.86 &$\checkmark$	&	&	&$\checkmark$	&	&	&$\checkmark$	&$\checkmark$ \\
                             & 1093.23 &$\checkmark$	&	&$\checkmark$	&$\checkmark$	&	&	&$\checkmark$	&$\checkmark$ \\
                             & 1093.40 &$\checkmark$	&$\checkmark$	&	&$\checkmark$	&	&	&$\checkmark$	&$\checkmark$ \\
  \hline
  Programme & log-PsML & Sex	& Region & Parents & School & Funding & Top 10\% & Pref. & Gap \\
 \hline
  \multirow{3}{*}{Chemistry} & -962.76 &  &  &  &  &  &  & $\checkmark$ & $\checkmark$ \\
                             & -963.77 &  &  &  &  &  & $\checkmark$ & $\checkmark$ & $\checkmark$ \\
                             & -963.81 &  &  & $\checkmark$ &  &  &  & $\checkmark$ & $\checkmark$ \\
  \hline
  Mathematics                & -1563.44 &$\checkmark$	&	&	&$\checkmark$	&	&$\checkmark$	&$\checkmark$	&$\checkmark$  \\
  and                        & -1564.27 &$\checkmark$	&	&$\checkmark$	&$\checkmark$	&	&$\checkmark$	&$\checkmark$	&$\checkmark$  \\
  Statistics                 & -1564.46 &$\checkmark$	&$\checkmark$	&	&$\checkmark$	&	&$\checkmark$	&$\checkmark$	&$\checkmark$  \\
  \hline
  \multirow{3}{*}{Physics}   & -550.78 &$\checkmark$	&	&	&$\checkmark$	&	&	&$\checkmark$	&$\checkmark$  \\
                             & -552.79 &$\checkmark$	&$\checkmark$	&	&$\checkmark$	&	&	&$\checkmark$	&$\checkmark$  \\
                             & -553.10 &$\checkmark$	&$\checkmark$	&	&	&	&	&$\checkmark$	&$\checkmark$  \\
  \hline
\end{tabular}
\end{table}

\begin{table}[h!]
\caption{\small Top 3 models with highest posterior probability (ticks indicate covariate inclusion).}\label{tableChapPUCProb} \centering \small \tabcolsep=0.11cm
\begin{tabular}[c]{clccccccccc}
  \hline
  Prior & Programme & Post. prob. & Sex	& Region & Parents & School & Funding & Top 10\% & Pref. & Gap \\
 \hline
  \multirow{9}{*}{(\ref{eqChapPUCunifprior1})} & \multirow{3}{*}{Chemistry} & 0.270 &	&$\checkmark$	&	&	&	&$\checkmark$	&$\checkmark$	&$\checkmark$ \\
  &                           & 0.238 &	&	&	&	&	&	&$\checkmark$	&$\checkmark$ \\
  &                           & 0.193 &	&$\checkmark$	&$\checkmark$	&	&	&	&$\checkmark$	&$\checkmark$ \\
  \cline{2-11}
  & Mathematics                & 0.942 &$\checkmark$	&	&$\checkmark$	&$\checkmark$	&$\checkmark$	&$\checkmark$	&$\checkmark$	&$\checkmark$ \\
  & and                        & 0.036 &$\checkmark$	&	&$\checkmark$	&$\checkmark$	&	&$\checkmark$	&$\checkmark$	&$\checkmark$ \\
  & Statistics                 & 0.014 &$\checkmark$	&$\checkmark$	&	&$\checkmark$	&	&$\checkmark$	&$\checkmark$	&$\checkmark$ \\
  \cline{2-11}
  &\multirow{3}{*}{Physics}   & 0.268 &	&	&	&	&	&	&	&  \\
  &                           & 0.150 &$\checkmark$	&	&$\checkmark$	&	&	&$\checkmark$	&$\checkmark$	&$\checkmark$  \\
  &                           & 0.054 &$\checkmark$	&	&	&$\checkmark$	&	&	&$\checkmark$	&$\checkmark$  \\
  \hline
  \multirow{9}{*}{(\ref{eqChapPUCunifprior2})} & \multirow{3}{*}{Chemistry} & 0.354 &	&	&	&	&	&	&	&  \\
  &                           & 0.259 &	&	&	&	&	&	&$\checkmark$	&$\checkmark$  \\
  &                           & 0.117 &	&$\checkmark$	&	&	&	&$\checkmark$	&$\checkmark$	&$\checkmark$  \\
  \cline{2-11}
  & Mathematics                & 0.982 &$\checkmark$	&	&$\checkmark$	&$\checkmark$	&$\checkmark$	&$\checkmark$	&$\checkmark$	&$\checkmark$ \\
  & and                        & 0.011 &$\checkmark$	&	&$\checkmark$	&$\checkmark$	&	&$\checkmark$	&$\checkmark$	&$\checkmark$ \\
  & Statistics                 & 0.004 &$\checkmark$	&$\checkmark$	&	&$\checkmark$	&	&$\checkmark$	&$\checkmark$	&$\checkmark$ \\
  \cline{2-11}
  & \multirow{3}{*}{Physics}   & 0.937 &	&	&	&	&	&	&	&  \\
  &                           & 0.009 &$\checkmark$	&	&$\checkmark$	&	&	&$\checkmark$	&$\checkmark$	&$\checkmark$  \\
  &                           & 0.007 &	&	&	&	&	&	&$\checkmark$	&$\checkmark$  \\
  \hline
\end{tabular}
\end{table}

Table \ref{tableChapPUCDICPsML} relates to Bayesian model comparison in terms of DIC and PsML. For the analyzed programmes, both criteria point in the same direction, suggesting that the most important covariates are the application preference and the gap indicator (associated with $\beta_{10}$ and $\beta_{11}$, respectively). Sex (related to $\beta_{1}$) and the high school type (represented by $\beta_{4}$ and $\beta_5$) are added to this list in case of Mathematics and Statistics students and the ones enrolled in Physics. The selection score indicator $\beta_9$ (top 10\%) also appears to have some relevance (specially for Mathematics and Statistics). As shown in Table \ref{tableChapPUCProb}, similar conclusions follow from the posterior distribution on the model space as those models with the highest posterior probabilities often include the same covariates  suggested by DIC and PsML. One difference is that for two programmes there is more support for the null model (the model without covariates where only the $\delta_{rt}$'s are included to model the baseline hazard). The choice between the priors in (\ref{eqChapPUCunifprior1}) and (\ref{eqChapPUCunifprior2}) on the model space can have a strong influence on posterior inference. As discussed in \cite{leysteel2009}, the prior in (\ref{eqChapPUCunifprior2}) downweighs models with size around $k^*/2=4$ (with priors odds in favour of the null model or the model with all 8 covariates versus a model with 4 covariates equal to 70) and this is accentuated in Physics, where the best model under both priors is the null model and the second best model has $k^*=5$, so that posterior model probabilities differ substantially between priors (see Table \ref{tableChapPUCProb}). In contrast, the choice between these priors has less effect in Maths and Stats, where the best models are of similar sizes. In a BMA framework, posterior probabilities of covariate inclusion are displayed in Table \ref{tableChapPUCProbs}. For these programmes, the highest posterior probabilities of inclusion relate to the application preference and the gap indicator (for both priors on the model space). As expected, results vary across programmes. For Mathematics and Statistics, there is strong evidence in favour of including all available covariates with the exception of the region of residence. In contrast, under both priors the model suggests that sex, high school type and the source of funding have no major influence on the academic outcomes of Chemistry students. For Physics (and to some extent for Chemistry) interesting models tend to be small and then the (locally) higher model size penalty implicit in prior (\ref{eqChapPUCunifprior2}) substantially reduces the inclusion probabilities of all covariates. For Maths and Stats, the best models are rather large and the prior (\ref{eqChapPUCunifprior2}) then favours models that are even larger, leading to very similar inclusion probabilities.

\begin{table}[h!]
\caption{\small Posterior probability of variable inclusion under priors (\ref{eqChapPUCunifprior1}) and (\ref{eqChapPUCunifprior2}) on the model space. }\label{tableChapPUCProbs} \centering \tabcolsep=0.11cm
\begin{tabular}[c]{llcccccccc}
  \hline
  Programme & Prior & Sex	& Region & Parents & School & Funding & Top 10\% & Pref. & Gap \\
  \hline
  \multirow{2}{*}{Chemistry} & (\ref{eqChapPUCunifprior1}) & 0.08 & 0.52 & 0.28 & 0.08 & 0.07 & 0.50 & 0.93 & 0.99 \\
                             & (\ref{eqChapPUCunifprior2}) & 0.06 & 0.23 & 0.15 & 0.04 & 0.06 & 0.27 & 0.61 & 0.65 \\
  \hline
  Maths. and  & (\ref{eqChapPUCunifprior1}) & 0.99 & 0.02 & 0.98 & 1.00 & 0.95 & 1.00 & 1.00 & 1.00 \\
  Statistics  & (\ref{eqChapPUCunifprior2}) & 1.00 & 0.01 & 0.99 & 1.00 & 0.98 & 1.00 & 1.00 & 1.00 \\
  \hline
  \multirow{2}{*}{Physics} & (\ref{eqChapPUCunifprior1}) & 0.49 & 0.25 & 0.37 & 0.31 & 0.11 & 0.27 & 0.71 & 0.62 \\
                           & (\ref{eqChapPUCunifprior2}) & 0.04 & 0.02 & 0.03 & 0.03 & 0.01 & 0.02 & 0.06 & 0.05 \\
  \hline
\end{tabular}
\end{table}

\begin{figure}
\begin{center}
\includegraphics[width=0.9\textwidth]{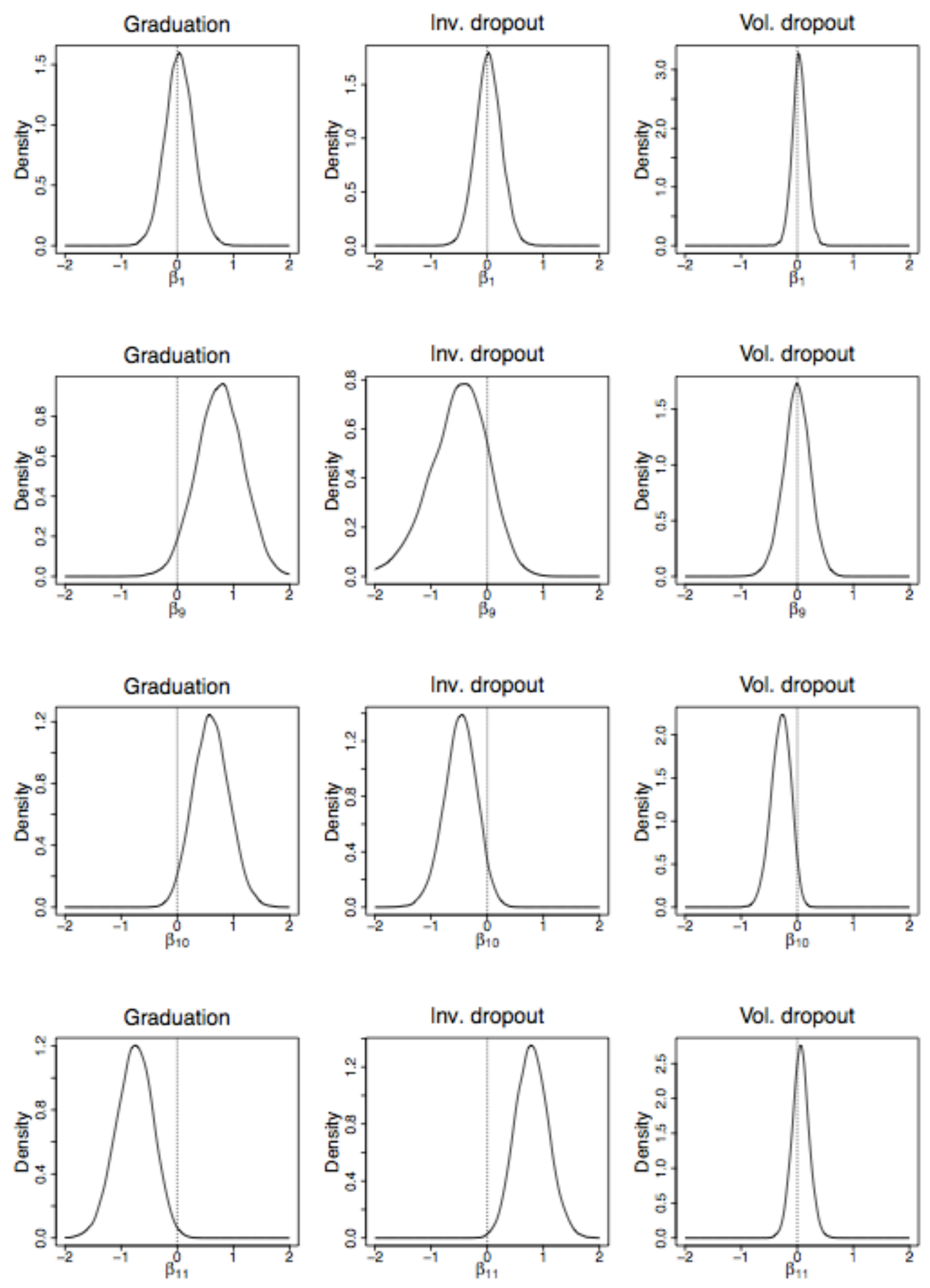}
\caption{\small Chemistry students: posterior density (given that the corresponding covariate is included in the model) of some selected regression coefficients: sex - female ($\beta_1$), selection score - top 10\% ($\beta_9$), preference - first ($\beta_{10}$) and gap - yes ($\beta_{11}$). A vertical dashed line was drawn at zero for reference. The prior in (\ref{eqChapPUCunifprior1}) was adopted.} \label{graphChapPUCPosterior}
\end{center}
\end{figure}

The posterior distribution of each $\beta_{rj}$ is given by a point mass at zero (equal to the probability of excluding the $j$-th covariate) and a continuous component (a mixture over the posterior distributions of $\beta_{rj}$ given each model where the corresponding covariate is included).
Figure \ref{graphChapPUCPosterior} displays the continuous component of the posterior distribution of some selected regression coefficients for the Chemistry programme under the prior in (\ref{eqChapPUCunifprior1}). The first row shows that the marginal densities of the effects related to sex are concentrated around zero. This is in line with the results in Table \ref{tableChapPUCProbs}, where both priors on the model space indicate a low posterior inclusion probability for sex. In contrast, the third row in Figure \ref{graphChapPUCPosterior} suggests a clear effect of the application preference on the three possible outcomes (positive for graduations and negative for both types of dropout). This agrees with a high posterior probability of inclusion and to put the magnitude of the effect into perspective, the odds for outcome $r=1,2,3$ versus no event are multiplied by a factor $\exp(\beta_{r~10})$ if Chemistry is the student's first preference. A similar situation is observed for the selection score indicator (see second row in Figure \ref{graphChapPUCPosterior}). In this case, those students with scores in the top 10\% graduate more and faster and are affected by less (and slower) involuntary dropouts. Nonetheless, this score indicator has no major influence on whether a student withdraws. Finally, for the gap indicator, we also notice a clear effect on graduations and involuntary dropouts, which has the opposite direction to that of the score indicator.

\section{CONCLUDING REMARKS} \label{ChapPUCSectionConcluding}	

The modelling of university outcomes (graduation or dropout) is not trivial. In this article, a simple but flexible competing risks survival model is employed for this purpose. This is based on the Proportional Odds model introduced in \cite{cox1972} and can be estimated by means of a multinomial logistic regression. The suggested sampling model has been previously employed in the context of university outcomes, but the structure of typical university outcome data precludes a maximum likelihood analysis. However, we use a Bayesian setting, where an appropriate prior distribution allows the extraction of sensible information from the data. Adopting a hierarchical structure allows for the derivation of a reasonably simple MCMC sampler for inference. The proposed methodology is applied to a dataset on undergraduate students enrolled in the Pontificia Universidad Cat\'olica de Chile (PUC) over the period 2000-2011.

As illustrated in Sections \ref{ChapPUCSectionDataset} and \ref{ChapPUCSectionAnalysis}, there are strong levels of heterogeneity between different programmes of the PUC. Hence, building a common model for the entire university is not recommended. For brevity, this article only presents the analysis of three science programmes for which late graduations and dropouts are a major issue, but the methodology presented here can be applied to all programmes. We formally consider model uncertainty in terms of the covariates included in the model. For the analyzed programmes, all the variable selection criteria (DIC, PsML and Bayes factors) tend to indicate similar results. However, in view of the posterior distribution on the model space, choosing a single model is not generally advisable and BMA provides more meaningful inference. The preference with which the student applied to the programme plays a major role in terms of the length of enrollment and its associated academic outcome for the three programmes under study. In addition, and perhaps surprisingly, having a gap between high school graduation and university admission is also found to be one of the most relevant covariates (but with the reverse effect of the preference indicator). The performance in the selection test is also generally an important determinant. Other factors, such as sex and the region of residence, only appear to matter for some of the programmes.

An obvious extension of the model presented here is to allow for different covariates in the modelling of the three  risks within the same programme. This would substantially increase the number of models in the model space, so we would need to base our inference on posterior model probabilities on sampling rather than complete enumeration. This can easily be implemented by extending the MCMC sampler to the model index and using \emph{e.g.}~Metropolis-Hastings updates based on data augmentation such as in \cite{holmesheld2006} or applications of the Automatic Generic sampler described by \cite{Green2003}.

\bibliographystyle{plainnat}
\bibliography{BibTeX_thesis}

\begin{thebibliography}{36}
\providecommand{\natexlab}[1]{#1}
\providecommand{\url}[1]{\texttt{#1}}
\expandafter\ifx\csname urlstyle\endcsname\relax
  \providecommand{\doi}[1]{doi: #1}\else
  \providecommand{\doi}{doi: \begingroup \urlstyle{rm}\Url}\fi

\bibitem[Albert and Anderson(1984)]{albertanderson1984}
A.~Albert and J.A. Anderson.
\newblock On the existence of maximum likelihood estimates in logistic
  regression models.
\newblock \emph{Biometrika}, 71:\penalty0 1--10, 1984.

\bibitem[Albert and Chib(1993)]{albertandchib1993}
J.~H. Albert and S.~Chib.
\newblock Bayesian analysis of binary and polychotomous response data.
\newblock \emph{Journal of the American Statistical Association}, 88:\penalty0
  669--679, 1993.

\bibitem[Arias~Ortis and Dehon(2011)]{ariasortiseanddehon2011}
E.~Arias~Ortis and C.~Dehon.
\newblock The roads to success: Analyzing dropout and degree completion at
  university.
\newblock Working Papers ECARES 2011-025, ULB - Universite Libre de Bruxelles,
  2011.

\bibitem[Bean(1980)]{bean1980}
J.P. Bean.
\newblock Dropouts and turnover: {T}he synthesis and test of a causal model of
  student attrition.
\newblock \emph{Research in Higher Education}, 12:\penalty0 155--187, 1980.

\bibitem[Chen et~al.(2008)Chen, Ibrahim, and Kim]{chenetal2008}
M.-H. Chen, J.G Ibrahim, and S.~Kim.
\newblock Properties and implementation of {J}effreys's prior in binomial
  regression models.
\newblock \emph{Journal of the American Statistical Association}, 103:\penalty0
  1659--1664, 2008.

\bibitem[Chipman et~al.(2001)Chipman, George, and McCulloch]{chipmanetal2001}
H.~Chipman, E.~I. George, and R.E. McCulloch.
\newblock \emph{The Practical Implementation of Bayesian Model Selection},
  volume~38 of \emph{Lecture Notes--Monograph Series}, pages 65--116.
\newblock Institute of Mathematical Statistics, Beachwood, OH, 2001.

\bibitem[Clerici et~al.(2014)Clerici, Giraldo, and
  Meggiolaro]{clericietall2013}
R.~Clerici, A.~Giraldo, and S.~Meggiolaro.
\newblock The determinants of academic outcomes in a competing risks approach:
  evidence from {I}taly.
\newblock \emph{Studies in Higher Education}, 2014.
\newblock URL \url{DOI: 10.1080/03075079.2013.878835}.

\bibitem[Cox(1972)]{cox1972}
D.R. Cox.
\newblock Regression models and life-tables.
\newblock \emph{Journal of the Royal Statistical Society. Series B},
  34:\penalty0 187--220, 1972.

\bibitem[Crowder(2001)]{crowder2001}
M.J. Crowder.
\newblock \emph{Classical competing risks}.
\newblock Chapman \& Hall/CRC, 2001.

\bibitem[Fern\'andez et~al.(2001)Fern\'andez, Ley, and
  Steel]{fernandezetal2001}
C.~Fern\'andez, E.~Ley, and M.F.J. Steel.
\newblock Model uncertainty in cross-country growth regressions.
\newblock \emph{Journal of Applied Econometrics}, 16:\penalty0 563--576, 2001.

\bibitem[Firth(1993)]{firth1993}
D.~Firth.
\newblock Bias reduction of maximum likelihood estimates.
\newblock \emph{Biometrika}, 80:\penalty0 27--38, 1993.

\bibitem[Forster(2010)]{forster2010}
J.J. Forster.
\newblock Bayesian inference for {P}oisson and multinomial log-linear models.
\newblock \emph{Statistical Methodology}, 7:\penalty0 210--224, 2010.

\bibitem[Fr{\"u}hwirth-Schnatter and Fr{\"u}hwirth(2010)]{fruhwirth2010}
S.~Fr{\"u}hwirth-Schnatter and R.~Fr{\"u}hwirth.
\newblock Data augmentation and {M}{C}{M}{C} for binary and multinomial logit
  models.
\newblock In T.~Kneib and G.~Tutz, editors, \emph{Statistical Modelling and
  Regression Structures}, pages 111--132. Springer, 2010.

\bibitem[Geisser and Eddy(1979)]{geisserandeddy1979}
S.~Geisser and W.F. Eddy.
\newblock A predictive approach to model selection.
\newblock \emph{Journal of the American Statistical Association}, 74:\penalty0
  153--160, 1979.

\bibitem[Gelman et~al.(2008)Gelman, Jakulin, Pittau, and Su]{gelmanetal2008}
A.~Gelman, A.~Jakulin, M.G. Pittau, and Y.S. Su.
\newblock A weakly informative default prior distribution for logistic and
  other regression models.
\newblock \emph{The Annals of Applied Statistics}, 2:\penalty0 1360--1383,
  2008.

\bibitem[Green(2003)]{Green2003}
P.J. Green.
\newblock Trans-dimensional {M}arkov chain {M}onte {C}arlo.
\newblock In P.J Green, N.L. Hjord, and S.~Richardson, editors, \emph{Highly
  Structured Stochastic Systems}, pages 179--198. Oxford University Press,
  2003.

\bibitem[Hanson et~al.(2014)Hanson, Branscum, and Johnson]{hansonetal2014}
T.E. Hanson, A.J. Branscum, and W.O. Johnson.
\newblock Informative $g$-priors for logistic regression.
\newblock \emph{Bayesian Analysis}, Forthcoming, 2014.

\bibitem[Hoeting et~al.(1999)Hoeting, Madigan, Raftery, and
  Volinsky]{hoetingetal1999}
J.A. Hoeting, D.~Madigan, A.E. Raftery, and C.T. Volinsky.
\newblock Bayesian model averaging: a tutorial.
\newblock \emph{Statistical Science}, 14:\penalty0 382--401, 1999.

\bibitem[Holmes and Held(2006)]{holmesheld2006}
C.C. Holmes and L.~Held.
\newblock Bayesian auxiliary variable models for binary and multinomial
  regression.
\newblock \emph{Bayesian Analysis}, 1:\penalty0 145--168, 2006.

\bibitem[Ibrahim and Laud(1991)]{ibrahimlaud1991}
J.G. Ibrahim and P.W. Laud.
\newblock On {B}ayesian analysis of generalized linear models using
  {J}effreys's prior.
\newblock \emph{Journal of the American Statistical Association}, 86:\penalty0
  981--986, 1991.

\bibitem[Ley and Steel(2009)]{leysteel2009}
E.~Ley and M.F.J. Steel.
\newblock On the effect of prior assumptions in {B}ayesian model averaging with
  applications to growth regression.
\newblock \emph{Journal of Applied Econometrics}, 24:\penalty0 651--674, 2009.

\bibitem[Ley and Steel(2012)]{leysteel2012}
E.~Ley and M.F.J. Steel.
\newblock Mixtures of $g$-priors for {B}ayesian model averaging with economic
  applications.
\newblock \emph{Journal of Econometrics}, 171:\penalty0 251--266, 2012.

\bibitem[Liang et~al.(2008)Liang, Paulo, Molina, Clyde, and
  Berger]{liangetal2008}
F.~Liang, R.~Paulo, G.~Molina, M.A. Clyde, and J.O. Berger.
\newblock Mixtures of $g$ priors for {B}ayesian variable selection.
\newblock \emph{Journal of the American Statistical Association}, 103:\penalty0
  410--423, 2008.

\bibitem[Murtaugh et~al.(1999)Murtaugh, Burns, and Schuster]{murtaughetat1999}
P.A. Murtaugh, L.D. Burns, and J.~Schuster.
\newblock Predicting the retention of university students.
\newblock \emph{Research in Higher Education}, 40:\penalty0 355--371, 1999.

\bibitem[Pintilie(2006)]{pintilie2006}
M.~Pintilie.
\newblock \emph{Competing Risks: A Practical Perspective}.
\newblock Statistics in Practice. Wiley, 2006.

\bibitem[Poirier(1994)]{poirier1994}
D.~Poirier.
\newblock Jeffreys' prior for logit models.
\newblock \emph{Journal of Econometrics}, 63:\penalty0 327--339, 1994.

\bibitem[Polson et~al.(2013)Polson, Scott, and Windle]{polsonetal2013}
N.~G. Polson, J.~G. Scott, and J.~Windle.
\newblock Bayesian inference for logistic models using {P}olya-{G}amma latent
  variables.
\newblock \emph{Journal of the American Statistical Association}, 108:\penalty0
  1339--1349, 2013.

\bibitem[Raftery et~al.(1997)Raftery, Madigan, and Hoeting]{rafteryetal1997}
A.E. Raftery, D.~Madigan, and J.A. Hoeting.
\newblock Bayesian model averaging for linear regression models.
\newblock \emph{Journal of the American Statistical Association}, 92:\penalty0
  179--191, 1997.

\bibitem[Roberts and Rosenthal(2009)]{robertsrosenthal2009}
G.O. Roberts and J.S. Rosenthal.
\newblock Examples of adaptive {M}{C}{M}{C}.
\newblock \emph{Journal of Computational and Graphical Statistics},
  18:\penalty0 349--367, 2009.

\bibitem[Scott and Kennedy(2005)]{scottkennedy2005}
M.A. Scott and B.B. Kennedy.
\newblock Pitfalls in pathways: {S}ome perspectives on competing risks event
  history analysis in education research.
\newblock \emph{Journal of Educational and Behavioral Statistics}, 30:\penalty0
  413--442, 2005.

\bibitem[Singer and Willett(1993)]{singerwillett1993}
J.D. Singer and J.B. Willett.
\newblock It's about time: {U}sing discrete-time survival analysis to study
  duration and the timing of events.
\newblock \emph{Journal of Educational and Behavioral Statistics}, 18:\penalty0
  155--195, 1993.

\bibitem[Singer and Willett(2003)]{singerwillett2003}
J.D. Singer and J.B. Willett.
\newblock \emph{Applied longitudinal data analysis: Modeling change and event
  occurrence}.
\newblock Oxford University Press, USA, 2003.

\bibitem[Spiegelhalter et~al.(2002)Spiegelhalter, Best, Carlin, and van~der
  Linde]{spiegelhalterelat2002}
D.J. Spiegelhalter, N.G. Best, B.P. Carlin, and A.~van~der Linde.
\newblock Bayesian measures of model complexity and fit (with discussion).
\newblock \emph{Journal of the Royal Statistical Society, B}, 64:\penalty0
  583--640, 2002.

\bibitem[Tinto(1975)]{tinto1975}
V.~Tinto.
\newblock Dropout from higher education: {A} theoretical synthesis of recent
  research.
\newblock \emph{Review of Educational Research}, 45:\penalty0 89--125, 1975.

\bibitem[Willett and Singer(1991)]{willettsinger1991}
J.~B. Willett and J.~D. Singer.
\newblock From whether to when: {N}ew methods for studying student dropout and
  teacher attrition.
\newblock \emph{Review of Educational Research}, 61:\penalty0 407--450, 1991.

\bibitem[Zellner(1986)]{zellner1986}
A.~Zellner.
\newblock On assessing prior distributions and {B}ayesian regression analysis
  with $g$-prior distributions.
\newblock In P.K. Goel and A.~Zellner, editors, \emph{Bayesian Inference and
  Decision Techniques: Essays in Honour of {B}runo de {F}inetti}, pages
  233--243, North-Holland: Amsterdam, 1986.

\end{thebibliography}

\end{document}